\documentclass[twocolumn,showpacs,preprintnumbers,amsmath,amssymb]{revtex4} 

\usepackage{graphicx,color}
\usepackage{dcolumn}
\usepackage{bm}

\begin{document}

\title{The unknotted strands of life: knots are very rare in RNA structures}

\author{C. Micheletti$^{1}$, M. Di Stefano$^{1}$ and H. Orland$^{2}$}
\affiliation{$^1$ SISSA - Scuola Internazionale Superiore di Studi Avanzati, Via Bonomea 265, 34136 Trieste (Italy)\\
$^2$ Institut de Physique Th\'eorique, CEA-Saclay, CEA 91191 Gif-sur-Yvette, France}

\date{October 4, 2014}

\begin{abstract}
  The ongoing effort to detect and characterize physical entanglement in
  biopolymers has so far established that knots are present in many
  globular proteins and also abound in viral DNA packaged inside
  bacteriophages. RNA molecules, on the other hand, have not yet been
  systematically screened for the occurrence of physical knots. We have
  accordingly undertaken the systematic profiling of the $\sim$6,000
  RNA structures present in the protein data bank. The search
  identified no more than three deeply-knotted RNA molecules. These are
  ribosomal RNAs solved by cryo-em and consist of about 3,000
  nucleotides.  Compared to the case of proteins and viral DNA, the
  observed incidence of RNA knots is therefore practically
  negligible. This suggests that either evolutionary selection, or
  thermodynamic and kinetic folding mechanisms act towards minimizing
  the entanglement of RNA to an extent that is unparalleled by other
  types of biomolecules. The properties of the three observed RNA
  knotting patterns provide valuable clues for designing RNA sequences 
  capable of self-tying in a twist-knot fold.
\end{abstract}

\maketitle

\section{Introduction}

The notion that biomolecules should be minimally entangled, so as to
fold efficiently and reproducibly and sustain functionally-oriented
structural arrangements, appears so intuitive that viable biological
molecules have long been thought to be free of physical
knots\cite{Virnau-Mallam-Jackson-knot-review:2011}.

Indeed, only several years after the publication of the human carbonic
anhydrase II structure~\cite{Eriksson:1988:Proteins:3151019} was it
realized that the enzyme backbone was tied in a trefoil
knot~\cite{Mansfield:1994:Nat-Struct-Biol:7656045}. More and more
instances of knotted and slipknotted proteins have since been
discovered\cite{Taylor:2000:Nature:10972297,King:2007:J-Mol-Biol:17764691},
and it is now established that about 2\% of the protein chains deposited
in the PDB host physical
knots~\cite{Potestio:2010:PLoS-Comput-Biol:20686683,Virnau-Mallam-Jackson-knot-review:2011}. The
latter range in complexity from the simplest trefoil knot to the six
crossing Stevedore's
knot~\cite{Bolinger:2010:PLoS-Comput-Biol:20369018}.

The functional implications of these knots are still
unclear. Nevertheless, several hypotheses have been formulated to
rationalize the functional advantage that arguably leads specific
knotted proteins to evolve from unknotted
ones~\cite{Potestio:2010:PLoS-Comput-Biol:20686683,Virnau:2006:PLoS-Comput-Biol:16978047}. For
instance, knots have been argued to enhance the mechanical stability of
active sites and preventing proteolytic
degradation~\cite{Virnau:2006:PLoS-Comput-Biol:16978047,Sulkowska2012}.

While empirical evidence demonstrates {\em de facto} that knotted
proteins do exist it also confirms the expectations that physical knots
can significantly hinder and slow down the spontaneous folding
process\cite{Mallam2012}. This is because the various folding steps need
to be well coordinated to ensure the formation of the correct knot type
in the correct protein
location\cite{Wallin2007,Sulkowska2009,aBeccara2013}.  This limited
kinetic accessibility is likely responsible for the much lower incidence
of knots in proteins compared to generic equilibrated
polymers~\cite{Lua:2006:PLoS-Comput-Biol:16710448}, where entanglement
inevitably arises with increasing chain length and
compactness~\cite{Sumners&Whittington:1988:J-Phys-A,Micheletti:2008:Biophys-J:18621819,Meluzzi:2010:Annu-Rev-Biophys:20192771,Micheletti20111}.

Physical knots have also been shown to occur abundantly in the
double-stranded DNA (dsDNA) of a number of
viruses~\cite{Liu:1981:Proc-Natl-Acad-Sci-U-S-A:6272306,Menissier11121984,Arsuaga:2002:Proc-Natl-Acad-Sci-U-S-A:11959991}. For
instance, a series of remarkable experiments carried out on the
bacteriophage P4
~\cite{Arsuaga:2002:Proc-Natl-Acad-Sci-U-S-A:11959991,Arsuaga:2005:Proc-Natl-Acad-Sci-U-S-A:15958528,Trigueros:2007:BMC-Biotechnol:18154674}
have shown that its 10kbp-long dsDNA contains many more physical knots
when packaged inside the phage capsid than when free in solution
\cite{Rybenkov:1993:Proc-Natl-Acad-Sci-U-S-A:8506378}. In fact, the P4
genome has a 95\% probability of accommodating complex physical knots
~\cite{Arsuaga:2002:Proc-Natl-Acad-Sci-U-S-A:11959991,Arsuaga:2005:Proc-Natl-Acad-Sci-U-S-A:15958528,Trigueros:2007:BMC-Biotechnol:18154674,Marenduzzo:2009:Proc-Natl-Acad-Sci-U-S-A:20018693}
when it is packaged inside a viral particle where it cannot be
simplified by topoisomerases\cite{Vologodskii2001,Liu2010}.

As for the case of proteins, the discovery of knots in packaged viral
DNA raised several questions about their functional implications,
particularly for the expected difficulty of ejecting the knotted genome
from the narrow capsid exit pore. More recent studies have shown that
this conundrum can be solved by considering the working of topological
friction at the molecular scale~\cite{Rosa:2012:Phys-Rev-Lett:23005684}
and especially the ordering effect of DNA self-interactions
\cite{Marenduzzo:2009:Proc-Natl-Acad-Sci-U-S-A:20018693} which favours
the untying of DNA knots inside the capsid during
ejection~\cite{Marenduzzo2013}.

Nowadays, the occurrence of physical entanglement in proteins and DNA,
is not only well-documented but is characterized well enough that novel
knotted proteins and short DNA molecules have been successfully designed
and the average entanglement of DNA filaments can be created or relaxed
in a controlled manner\cite{King2010,Shih2010,Tang27092011}.

These topological profiling efforts, however, have not been paralleled
for the third and last kind of strand of
life\cite{KnottedStrandsOfLife}, namely RNA. To the best of our
knowledge no systematic survey of physical knots in RNAs has been
carried out so far and no physical knots have been reported in
naturally-occurring RNA structures.

The interest in characterizing and detecting other forms of RNA
entanglement has nevertheless been significant in past years,
particularly regarding pseudo-knots. These are secondary structure
elements with a non-trivial geometry which abound in RNA molecules and
which can have important functional
implications~\cite{Rietveld1982,Rietveld1983,Giedroc2000,Cho13102009}. In
particular, several efforts are being spent to clarify how exactly
pseudoknots can hinder RNA translocation through the ribosomal pore and
cause a shift in the codon reading
frame~\cite{Giedroc2000,Chen:2009:PNAS,Hansen:2009:PNAS,Ritchie:2012:PNAS}.

\begin{figure}[tbp!]
\centering
\vspace*{.05in}
\includegraphics[width=0.75\columnwidth]{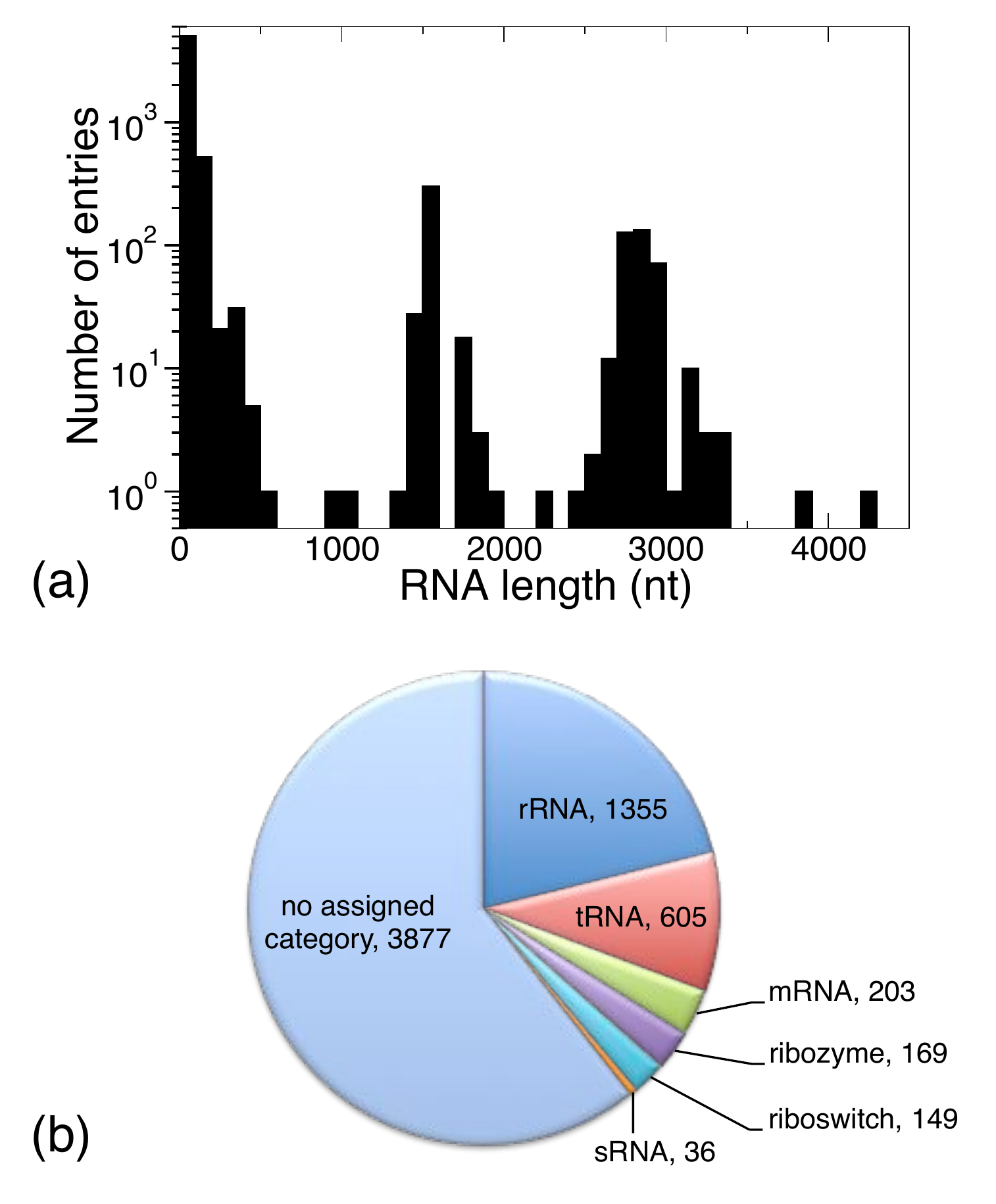}
\caption{RNA dataset properties. The length distribution and categorized
  subdivision of the considered RNA chains are given in panels (a) and
  (b), respectively. The category assignment is based on the COMPND field of the
the PDB structure files.}
\label{fig:Distributions}
\end{figure}

Qualitatively however, pseudoknots are very different from physical
knots. In fact, as their name suggest, they are not real knots: by
pulling on the ends of a chain, any pseudoknot will progressively yield
until the chain is fully stretched, while physical knots will never
disappear but rather tighten up.

\begin{table*}[btp!]
  \begin{tabular*}{\hsize}{@{\extracolsep{\fill}}lllclll}Knot&Molecule&Organism&\# nucleotides&PDB ID& Essential knotted region&Essential crossings\cr
    \hline
\cr
$4_1$       &23S rRNA&\emph{Escherichia}&2,740&2GYA:0&A1434--U1563               &G1478--C1480 and C1558--C1561 \cr
            &        &\emph{coli}       &     &      &                           &                              \cr
\cr
$16_{124834}$&26S rRNA&\emph{Thermomyces}&3,170&3JYX:5&A416--A428, G616--C636,    &C618--A622 and A1401--U1405   \cr 
            &        &\emph{lanuginosus}&     &      &A647--A677, C700--G716,    &A710--G714 and U2775--A2780   \cr
            &        &                  &     &      &G787--A791, A1373--A1433,  &                              \cr
            &        &                  &     &      &U1439--U1448, C2389--C2405,&                              \cr 
            &        &                  &     &      &G2619--C2625, C2760--U2822,&                              \cr
            &        &                  &     &      &U2978--U2999               &                              \cr 
\cr
$3_1$\#$3_1$\#$4_1$\#$3_1$       &23S rRNA&\emph{Escherichia}&2,904&1C2W:B&G254-C366    \phantom{xx}  ($3_1$)               &G263--C264 and G363--C364     \cr
       &        &\emph{coli}       &     &      &G520-A825 \phantom{xx}   ($3_1$)                &C581--G583 and C814--C816     \cr
       &        &                  &     &      &U1440-A1535 \phantom{xx}   ($4_1$)                &C1454--G1455 and C1526--G1527 \cr
       &        &                  &     &      &                           &U1476--A1477 and G1514--A1515 \cr
       &        &                  &     &      &U1851-C1892    \phantom{xx}   ($3_1$)             &U1856--G1857 and C1887--G1888 \cr
    \hline
  \end{tabular*}
\label{tab:knots}
\caption{{\bf Knotted RNA structures} For the 23S knotted units the
  knotted regions correspond to relatively short, uninterrupted
  stretches of the backbone which accommodate the whole knot (for
  2GYA:0) or its separable prime components (for 1C2W:B), see Figs.~3
  and 4.  To expose with maximum clarity the more delocalised knot of
  the 26S unit 3JYX:5, we list the sequence of backbone segments that,
  once joined, embody the fundamentally-entangled portion of this
  molecule, see Fig.~2. Such essential knotted region was obtained by
  ``shortcutting'' nugatory helices and loops that do not contribute to
  the chain entanglement. The essential crossings consist of pairs of
  facing RNA strands which can untie or simplify the knot when passed
  through each other. The nucleotide numbering follows the indexing of
  the associated PDB file.}
\end{table*}

Clearly, the impact of knots on RNA mechanical resistance, and hindrance
to translocation can be much more dramatic than that of pseudoknots, and
hence establishing their degree of occurrence can shed light on their
biological relevance and implications.

In this respect it is important to recall the seminal study of Wang {\em
  et al.}~\cite{Wang03091996}, who succeeded in designing an RNA
sequence of 104 nucleotides capable of folding into a knotted
structure. The study not only gave a proof of concept that RNA can be
knotted, but also established that the {\em Escherichia coli} DNA
topoisomerase III could catalyze the interconversion of knotted and
unknotted forms of the synthetic RNA. Clearly, the very existence of
enzymes acting as RNA topoisomerases points to their putative role in
simplifying RNA entanglement, including knots, {\em in vivo}.  As a
matter of fact, the recently discovered eukaryotic circular RNAs
(circRNAs)~\cite{10.1371/journal.pone.0030733,Memczak2013,Jeck01022013,10.1371/journal.pgen.1003777}
have been argued to be good candidates for hosting knots which could be
permanently trapped by their circular
structure~\cite{Frank-Kamenetskii13}.

\section{Results and Discussion}

These considerations motivated us to carry out a systematic search for
the occurrence of physical knots on all the RNA structures that are
presently available in the protein data bank (PDB)\cite{Berman2000}.

To this purpose we considered all currently available PDB entries and
isolated 2,863 of them which involve RNA either alone or in association
with proteins and DNA or hybridized with DNA. The total number of
distinct RNA chains extracted from these entries was equal to 6,394. The
length distribution of the considered RNA chains and their categorized
subdivision is given in Fig.~1.

The backbone (P-atoms trace) of each RNA chain was circularized ``in
silico'' using the so-called minimally-interfering closure
scheme\cite{Min_entang_closure}. This algorithm turns the linear, open,
backbone of the molecule into a closed structure which hence has a
mathematically well-defined topological state. The knotted state of the
closed backbone was finally established by computing its Dowker code
which is a non-ambiguous topological indicator that can be used to
identify prime knots of up to 16 crossings by comparison against
available look-up tables, see Methods.

In our systematic survey we found only three instances of RNA
molecules that accommodate physical knots in their structurally-intact
(no breaks) backbone. The properties of these knotted structures are
given in Table~1, where they are listed in order of increasing
topological complexity and chain length.

The first instance corresponds to the figure-of-eight or Flemish knot,
$4_1$, and is found in the {\em Escherichia coli} 23S ribosomal RNA. The
second instance is a much more complex 16-crossings prime knot which is
present in the {\em Thermomyces lanuginosus} 26S ribosomal unit.
The two molecules consist of about $2,800$ and $3,200$ nucleotides,
respectively.  Finally, the third knotted structure is, again, an {\em
  Escherichia coli} 23S ribosomal RNA. Unlike the first instance,
however, it features a composite knot formed by the succession of four separate
prime knotted components: three trefoil knots plus a figure-of-eight
knot.

\begin{figure*}[bpt!]
\vspace*{.05in}
\centering
\includegraphics[width=0.9\textwidth]{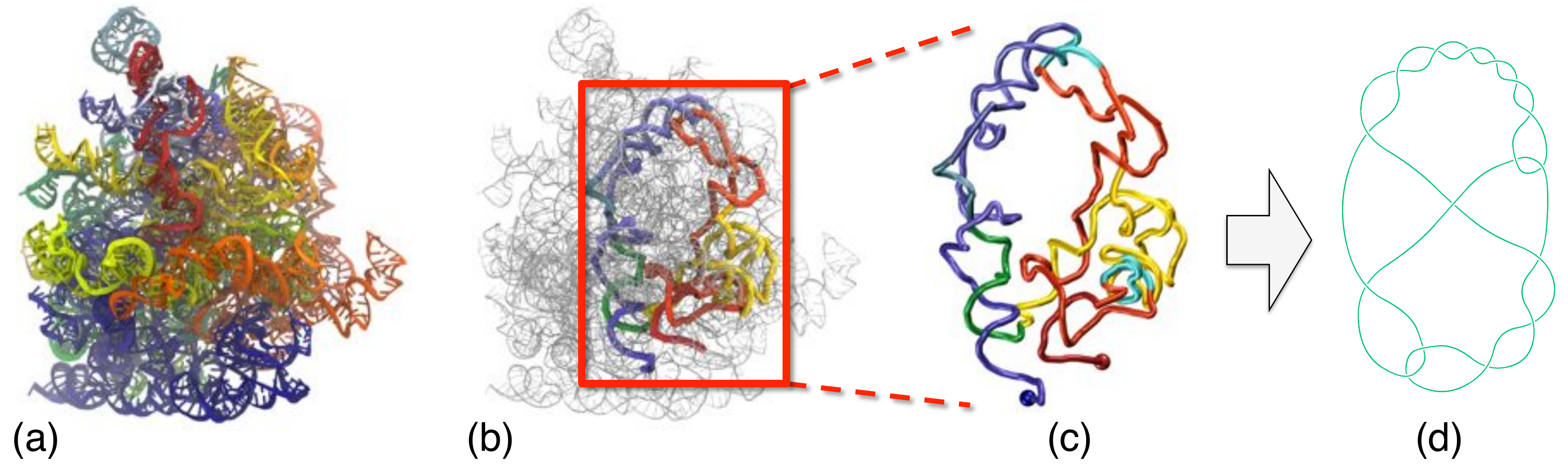}
\caption{Knotted 26S ribosomal RNA structure from PDB entry 3JYX:5. The
  complete structural representation of the RNA chain is given
  in panel (a) and is colored according to a rainbow scheme, red $\to$
  yellow $\to$ green $\to$ blue, across the two termini. The essential
  knotted region, tied in a 16-crossings knot, is highlighted in panel
  (b) and is presented in excerpted form in panel (c). The short cyan
  regions in panel (c) highlight the essential crossings of the knot,
  see Table~1. The minimal ring diagram of the corresponding
  16-crossings knot is shown in panel (d).
\label{Fig:3JYX5}}
\end{figure*}

We start by discussing the complex knot found for the 26S ribosomal unit,
which is shown in Fig.~2. In this figure, panel (a)
represents the whole molecule, while the ``essential'' knotted region is
shown in panels (b) and (c). This essential region was obtained by
shortcutting helices that do not contribute to the topological
entanglement of the molecule, so as to make its knotted state readily
perceivable by visual inspection.  Panel (d), instead, is a minimal
diagrammatic representation of the closed physical knot.  As it is
apparent from panels (c) and (d), the high nominal complexity of this
entanglement, which is measured through the number of crossings in the
simplest diagrammatic projection, mostly arises because of the clasp
formed by two helices located at the top of panel (c).  The nucleotides
at the clasp have a  large sequence separation, about 2,000
nucleotides, which underscores the significant non-locality of the
knot. It is important to notice that the knot would persist even if this
clasp was removed by a suitable strand passage. In fact, in this case,
the original sixteen-crossings knot would simplify to a seven-crossing
one ($7_2$) which can finally be untied by a further virtual strand
passage. The two regions which are arguably naturally primed for such
simplifying strand-passages are listed in Table~1 and highlighted in
cyan in panel (c).

\begin{figure*}[tbp!]
\centering
\vspace*{.05in}
\includegraphics[width=0.9\textwidth]{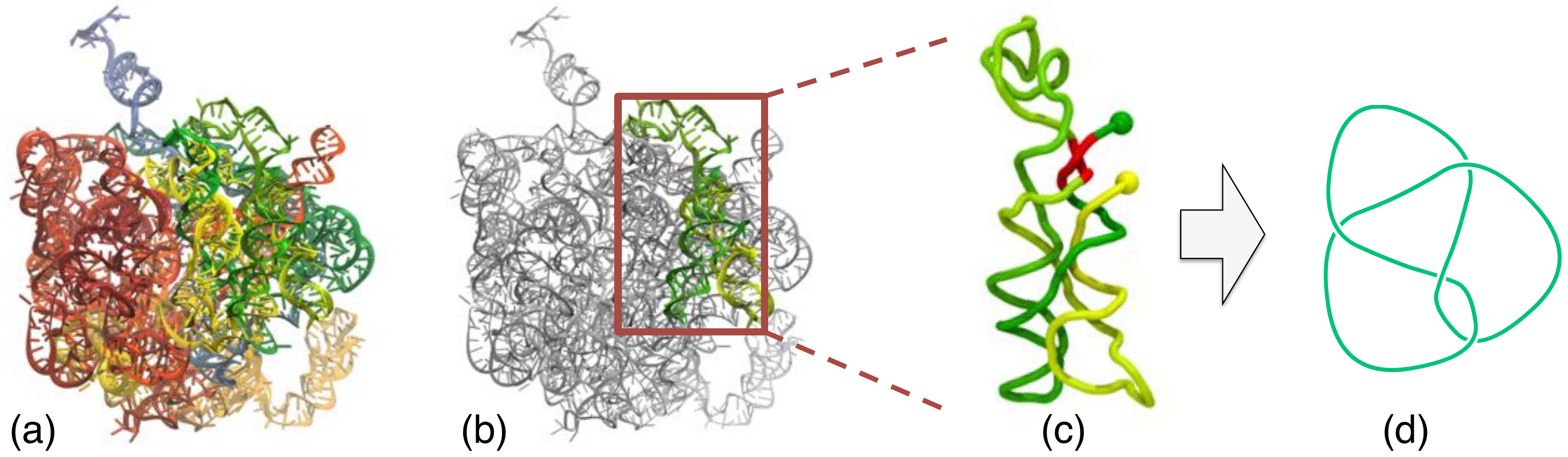}
\caption{Knotted 23S ribosomal RNA structure from PDB entry 2GYA:0. The
  complete structural representation of the ribosomal structure is given
  in panel (a) and is colored according to a rainbow scheme across the
  two termini. The knotted region, tied in a figure-of-eight or $4_1$
  knot, is highlighted in panel (b) and is presented in excerpted form
  in panel (c). The red regions in panel (c) highlight the essential
  crossing of the knot, see Table~1. The minimal ring diagram of the
  figure-of-eight knot is shown in panel (d).
\label{Fig:2GYA0}}
\end{figure*}

We now turn to the two knotted 23S units from {\em Escherichia coli},
starting from chain 2GYA:0 which is shown in Fig.~3a.  The
entangled region of this molecule spans as few as 150 nucleotides and
its knotted state, corresponding to a simple figure-of-eight ($4_1$)
knot, is clearly seen in the excerpted knotted region of Fig.~3c.  The featured
figure-of-eight knot is an achiral twist knot. These kind of knots are
easily produced by a single strand passage in rings that are repeatedly
twisted. It is intriguing to notice: (i) the analogy of this mechanism
with the strand passages occurring in RNA helices as in the previous
case and (ii) this mechanism is utterly different from the one emerging
in densely packed double-stranded DNA, where torus knots are abundant
and twist knots
rare\cite{Arsuaga:2005:Proc-Natl-Acad-Sci-U-S-A:15958528,Trigueros:2007:BMC-Biotechnol:18154674}. Unlike
the discussed 26S case, a single strand passage suffices to untie the
observed $4_1$ knot. A possible essential crossing, where the untying
strand passage could be performed, is listed in Table~1
and highlighted in red in Fig.~3c.

The other knotted instance of the 23S unit, corresponding to the 1C2W:B
chain, is finally shown in Fig.~4. It features a composite
knot resulting from the concatenation of four separate prime components:
a figure-of-eight knot and three trefoil ones, see
Fig.~4b. The entanglement of the shortest component, which
consists of only $\sim40$ nucleotides is clearly visible in the backbone trace
of Fig.~4c.

In connection with RNA self-entanglement, it is worth recalling that it has
been proposed to classify RNA structures by their topological
genus~\cite{Orland_Zee,Bon2008}. The genus provides a useful
characterization of the complexity of pseudoknots and can be used for
secondary structure prediction~\cite{Bon2011,Bon2013}. It is thus
interesting to ascertain if it correlates with the nominal complexity of
the knots found in the RNA. Accordingly, we used the RNApdbee web
server~\cite{Antczak2014}, to extract the secondary structures of
the RNA from their PDB files and then computed the corresponding genii
using the computational engine of the McGenus web server
\cite{Bon2013}. We found that the two 23S ribosomal units, 2GYA:0 and
1C2W:B, have genus respectively equal to $g=6$ and $g=7$, while the
knotted 26S ribosomal unit has genus equal to $g=5$. As was shown
in~\cite{Bon2008}, these genii are fairly small, but still compatible with the
typical size of $3,000$ nucleotides of these RNAs.

The fact that only three RNA molecules are knotted over a set of about
$6,000$ entries indicates the extreme paucity of non-trivial
entanglement in naturally-occurring RNAs. As a matter of fact, the knots
incidence is so low that one may doubt whether the three exceptional
structures are genuinely knotted.

This point is particularly pertinent because
both the 23S and 26S ribosomal RNAs were solved by
cryo-electron microscopy (cryo-em). This technique has
proved invaluable for gaining quantitative insight into the structural
organization of large and complex biomolecular structures, including the
three knotted instances discussed here. At the same time, its scope can
be limited in practice by two main factors. First, the electronic flux
impacting the molecules may be high enough to alter their
structures. Secondly, the resolution of cryo-em maps is appreciably
lower than in conventional crystallography and hence is prone to
ambiguous model reconstruction without suitable knowledge-based
constraints. For a better control of the latter ambiguities, several
order parameters are usually monitored to establish the local quality of
the model fit of the electron density map.

\begin{figure}[bp!]
\centering
\vspace*{.05in}
\includegraphics[width=\columnwidth]{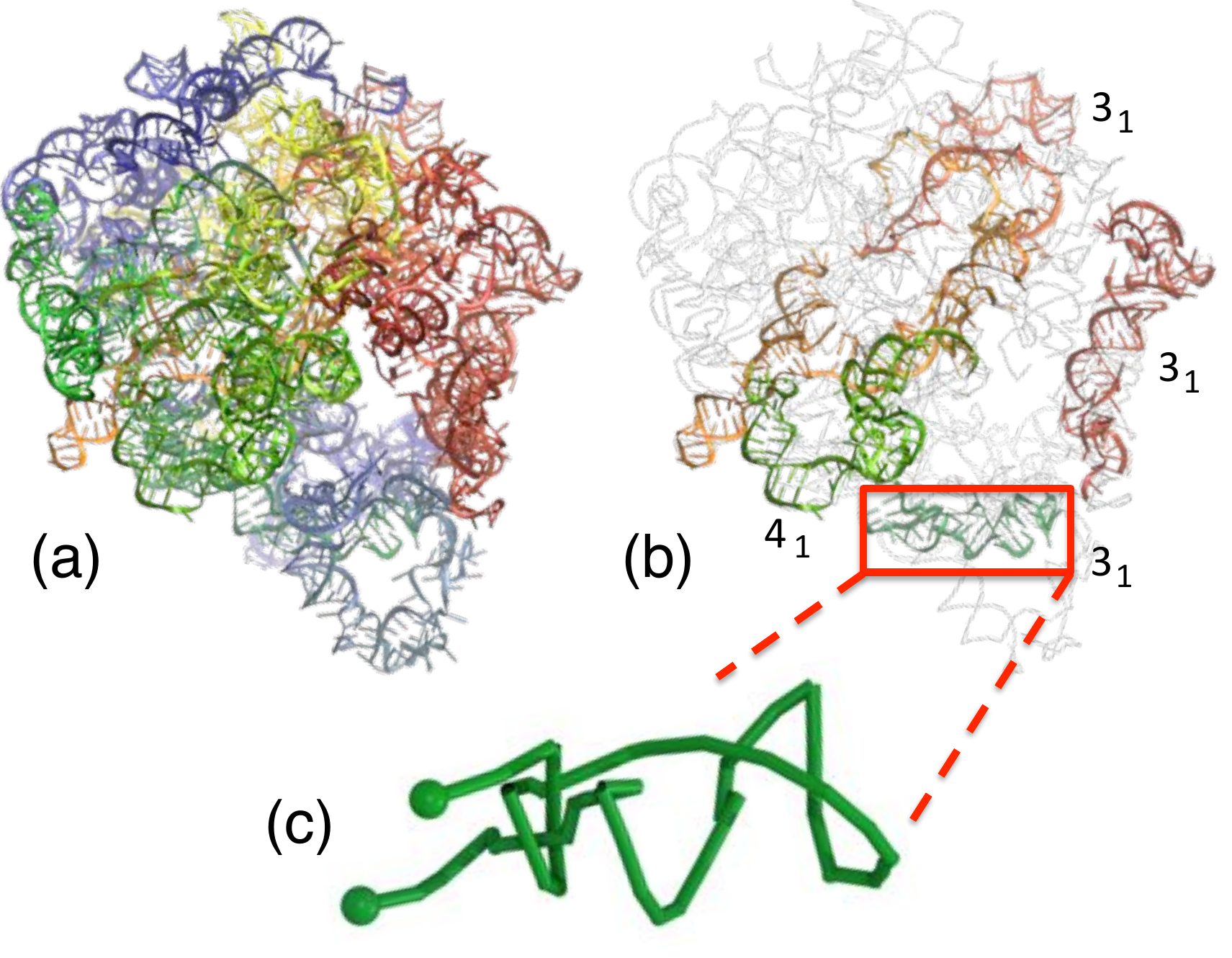}
\caption{Knotted 23S ribosomal RNA structure from PDB entry 1C2W:B. The
  complete structural representation of the RNA chain is given in panel
  (a) and is colored according to a rainbow scheme across the two
  termini. The four prime components of the knotted region are
  highlighted in color in panel (b). The entanglement of the shortest
  trefoil-knotted component (U1851--C1892) is shown in panel (c). \label{Fig:1C2WB}}
\end{figure}

Such quality parameters are available for PDB entries 2GYA:0 and 3JYX:5.
In both cases, it is found that the regions corresponding to the knots'
``essential crossings'' fall within, or close to, the range of
nucleotides where the model fitting was marked as deviating appreciably
from the target electron density map. It is therefore plausible that the unknotted
structure obtained by eliminating the essential crossings could provide
a better fit to the target em map.  This could, in turn, imply that none
of the thousands of available RNA structures are knotted.

Further elements in favour of this conclusion arise when comparing the
knotted RNAs with other related 23S and 26S units available in the PDB.
In particular, besides 2GYA:0 and 1C2W:B there exist 36 other PDB
entries of the {\em Escherichia coli} 23S unit. Strikingly, these
alternative structures are all unknotted. They include the 2I2T:B and
2I2V:B chains solved by X-ray at a 3.2\AA\ resolution and the 2GYC:0
cryo-em structure which was solved with same technique used for the
knotted 2GYA:0 counterpart, of which it shares the sequence and length.
In particular we note that, by applying the ARTS structural alignment\cite{Dror01072006}
algorithm to 2GYA:0 and 2GYC:0 it is found that the two strands
1558-1561 and 1478-1480 which form the essential clasp shown in Fig.~3c
for the knotted 2GYA:0 chain, are not interlocked in the unknotted
2GYC:0 structure.  It would therefore be interesting to
ascertain if the cryo-em data of 2GYA:0 could be better fitted by
removing the clasp.

For the 26S ribosomal unit, instead, there exist only two structures in
the PDB, including the previously-discussed knotted one for {\em
  Thermomyces lanuginosus}. The second structure is for {\em Tetrahymena
  thermophila} and is unknotted. This structure was solved by X-ray at a
resolution of 3.5\AA\ and its asymmetric unit is split over PDB entries
4A1D:1, 4A18:1, 4A19:1.  The two 26S structures, despite differing by
knotted state, have very similar sequences. Their mutual sequence
identity from a BLASTn alignment\cite{Altschul1997} is, in fact,
82\%. The extensive sequence alignment includes a region encompassing
one of the two essential crossings for the knotted 26S structure of {\em
  Thermomyces lanuginosus}. The two involved RNA segments, which are
interlocked in the {\em Thermomyces lanuginosus} 26S unit , are unlocked
in the {\em Tetrahymena thermophila} one. Again, towards
clarifying the genuine character of the entanglement of the 26S unit knot it could be verified whether
removing the interlocking of the two segments improves the cryo-em
data fitting.

Regardless of the fact that the pool of knotted RNA entries consists of
only three or zero entries, the outcome of the present survey is that the
incidence of knots in RNA molecules is utterly negligible both in
absolute terms and also by comparison with proteins and viral dsDNA.

Therefore, within the realm of the various ``strands of life'', RNA seems to be the only instance where physical knots
are most rare, and possibly absent altogether.

The striking observation that RNA structures are virtually free of 
physical knots pose the question of understanding what
plausible mechanisms may have ruled out the occurrence of any major type
of entanglement in RNA. {\em A priori} it is possible to envisage
several possible selection processes that involve either the kinetics or the
thermodynamics of RNA folding.

On the one hand, it is known that the fold organization of
naturally-occurring RNA sequences is much simpler than for random sequences
with the same overall nucleotide composition. This point is well-illustrated by
considering the above-mentioned genus as an indicator of the complexity
and entanglement of RNA secondary structures. In fact, one observes that
the minimum energy structures of random sequences of $L=3,000$
nucleotides typically have a genus of the order of $0.13 L = 390$
\cite{Ribeca} while naturally-occurring ones of the same length have a
genus of only 5 to 8.  This provides a strong evidence that
naturally-occurring RNA sequences have arguably evolved to minimize
their geometrical complexity and hence the entanglement of their native
folds.

On the other hand, this sequence-encoded simplicity of RNA structures
may be further aided by the kinetics of the folding process. In
fact, it may be envisaged that the folding of long RNA molecules may
occur, at least in part, co-transcriptionally. This mechanism ought to
favor the formation of local helices in newly transcribed regions. In
this case, it would be very difficult to develop knots in the resulting
highly-branched structure of long and helically-folded RNA.

\begin{figure}[bp]
\centering
\includegraphics[width=0.35\columnwidth]{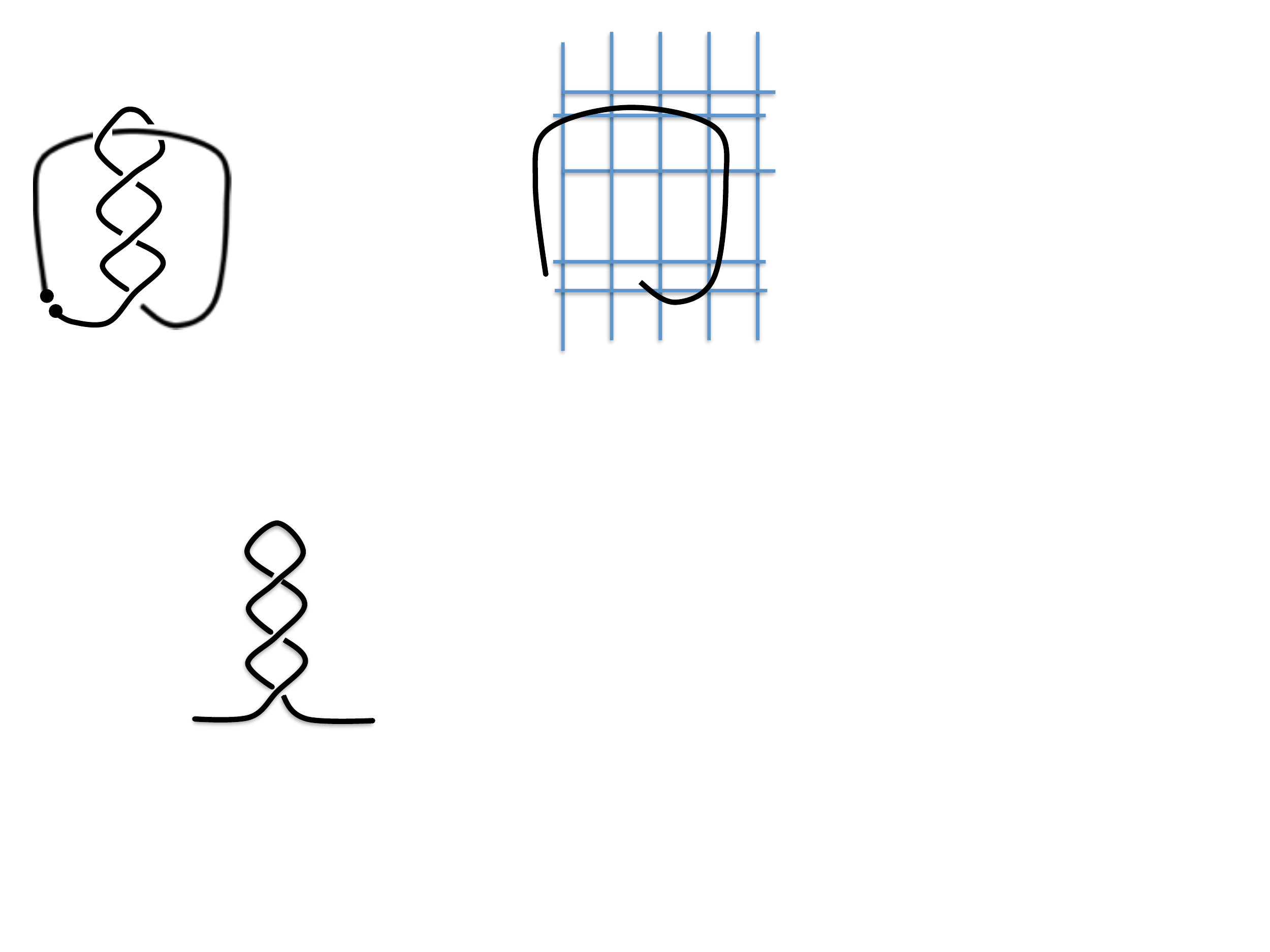}
\caption{Design of RNA twist knots. Twist knots (such as the $5_2$ knot shown here) can be formed by RNA sequences designed to fold into a helix with an unpaired loop large enough to be threaded by one of the two termini. The knot could be stabilised by base pairing at the clasp (helix apex) or by annealing of the two complementary termini.}
\label{fig:twistknots}
\end{figure}

Further clues regarding the relative weight of these two, possibly
concurrent mechanisms as well as the functional implications of RNA
knots could be provided in the future by the investigation of the
newly-discovered eukaryotic circRNAs, where non-trivial structural
entanglement is expected to arise from their circularization and could
possibly be co-opted to hinder their {\em in vivo}
degradation\cite{10.1371/journal.pone.0030733,Memczak2013,Jeck01022013,10.1371/journal.pgen.1003777,Frank-Kamenetskii13}.

Finally, we note that the structural properties of the three knotted
RNAs suggest how to generalize the strategy previously followed by
Seeman and coworkers\cite{Seeman97} to design novel RNA sequences
capable of self-tying into a fold with $3_1$ and $4_1$ knots topology.
In fact, a possible scheme would be to promote first the formation of a
helix and then the threading of its apical loop, as sketched in
Fig.~5. As a matter of fact, by increasing the length of the stem one
could generate molecular knots of far greater complexity than the
simplest prime knots achieved so far. This is because by modulating the
number of turns in the helix it would be possible to span across the
various members of the twist knots, family. The helix length modulation,
could be achieved either by increasing the length of the involved
sequence, or even by varying the concentration and type of counterions
in solution that could affect both the geometry of the
helix\cite{Seeman97} and the electrostatic persistence length
controlling the knot size\cite{Dommersnes:PRE:2002}.

\phantom{XXXXX}

\section{Methods}
\subsection{RNA structures database}

 The processed database of RNA chains was
obtained by selecting all protein data bank (PDB) entries that
contained RNA molecules either alone or complexed/hybridized with other
types of biomolecules such as proteins and DNA. Applying such selection
criterion to the structures available as of June 2014 returned 2,863 PDB
entries, each comprising one or more RNA chains. Of these chains we
retained only those that contained RNA nucleotides. Following the
criteria adopted by the Jena Library of Biological Macromolecules
http://jenalib.fli-leibniz.de/IMAGE.html, this automatic filtering was
performed by checking for the presence of at least one nucleotide
O2$^\prime$ atom or of uracil-related residue identifier. Nucleotide
entries corresponding to alternate locations in the PDB file are
neglected. The resulting structural dataset consisted of 6,394 RNA chains.

\subsection{Detecting and locating knots in RNA chains.}  

To detect a physical
knot in an RNA chain we first bridge the two termini of its backbone
(phosphate trace) with a minimally-invasive closing arc
\cite{Min_entang_closure} so as to obtain a ring.  The knotted state of
the ring is then established by computing the Alexander determinants and
the Dowker code after a topology-preserving simplification of the ring
geometry
\cite{KoniarisMuthukumar:1991a,Taylor:2000:Nature:10972297,Micheletti20111}.
Applying the topological-profiling scheme to the 6,394 RNA chains
returned only nine putatively knotted structures which were further
screened for structural integrity. Specifically, to rule out significant
structural gaps, we disregarded those chains where consecutive
phosphates are farther apart than 15\AA. Only the three structures
listed in Table~1 were found to have a knotted and gapless backbone. The
portions of these chains that accommodate the physical knots (or its
separable prime components) were located by using the so-called bottom-up
knot search scheme which identifyies the shortest stretch of the
phosphate trace that, after closure, has the same desired topology.
chain \cite{Tubiana2011}. 

\section{acknowledgments} 

We thank Giovanni Bussi and Stefano Gustincich for extensive
discussions.  This work was partially supported by the Italian Ministry
of Education grant PRIN No.~2010HXAW77.


\end{document}